\documentclass[conference]{IEEEtran}
\IEEEoverridecommandlockouts
\usepackage{cite}
\usepackage{comment}
\usepackage{amsmath,amssymb,amsfonts}
\usepackage[inkscapelatex=false]{svg}
\usepackage[inline]{enumitem}
\usepackage{algorithmic}
\usepackage{graphicx}
\usepackage{textcomp}
\usepackage{listings}
\usepackage{adjustbox}

\usepackage[hyphens]{url}
\usepackage[hidelinks]{hyperref}
\hypersetup{breaklinks=true}

\usepackage{xcolor}
\lstdefinelanguage{json}{
    basicstyle=\ttfamily\scriptsize, % Use typewriter font, adjust size as needed
    lineskip=-2pt,
    numbers=left,                      % Line numbers on the left
    numberstyle=\tiny\color{gray},     % Line numbers style
    stepnumber=1,                      % Number every line
    numbersep=5pt,                     % Space between line numbers and code
    showstringspaces=false,            % Do not show spaces in strings
    captionpos=t,
    backgroundcolor=\color[gray]{0.95},% Light gray background
    stringstyle=\color{red},           % Strings in red
    commentstyle=\color{green},        % Comments in green (if any)
    keywordstyle=\color{blue}\bfseries,% Keywords in blue bold
    morestring=[b]",
    morecomment=[l]{//},
    morekeywords={true, false, null}   % JSON-specific keywords
}
\begin{document}
\bstctlcite{IEEEexample:BSTcontrol}

\title{CAMINO: Cloud-native Autonomous Management and Intent-based Orchestrator

% \thanks{This work is a contribution by Project REASON, a UK Government funded project under the Future Open Networks Research Challenge (FONRC) sponsored by the Department of Science Innovation and Technology (DSIT).}
}

\author{\IEEEauthorblockN{Konstantinos Antonakoglou, Ioannis Mavromatis, Saptarshi Ghosh, Mark Rouse, Konstantinos Katsaros}
\IEEEauthorblockA{\textit{Digital Catapult, London, UK} \\
Emails: \{konstantinos.antonakoglou, ioannis.mavromatis, saptarshi.ghosh, mark.rouse, kostas.katsaros\}@digicatapult.org.uk}
}

\maketitle

\begin{abstract}
This paper introduces CAMINO, a Cloud-native Autonomous Management and Intent-based Orchestrator designed to address the challenges of scalable, declarative, and cloud-native service management and orchestration. CAMINO leverages a modular architecture, the Configuration-as-Data (CaD) paradigm, and real-time resource monitoring to facilitate zero-touch provisioning across multi-edge infrastructure. By incorporating intent-driven orchestration and observability capabilities, CAMINO enables automated lifecycle management of network functions, ensuring optimized resource utilisation. The proposed solution abstracts complex configurations into high-level intents, offering a scalable approach to orchestrating services in distributed cloud-native infrastructures. This paper details CAMINO's system architecture, implementation, and key benefits, highlighting its effectiveness in cloud-native telecommunications environments.
\end{abstract}


\vspace{-2mm}
\section{Introduction}

The cloud computing and telecommunications industry stakeholders face increasingly complex and multifaceted challenges in the management and orchestration (MANO) of service and network intents, particularly as deployments become more heterogeneous in terms of underlying technologies and system scale~\cite{9302614}. The expectations for autonomous systems with automation capabilities are growing, necessitating closed-loop control at different levels and phases of MANO workflows.

To enable such capabilities, Network Function Virtualisation (NFV) MANO platforms allow services to consume computing and network resources (both physical and virtual) offered by infrastructure providers (InPs). These resources are distributed across distinct administrative domains (e.g., belonging to network operators) to support the deployment of interconnected meshes of Virtual Network Functions (VNFs) and Cloud-native Network Functions (CNFs).

According to ETSI Zero-touch Network and Service Management (ZSM) specifications~\cite{etsi_gr_zsm_011_v2_1_1}, a management or administrative domain is an autonomous entity that groups computing and network resources under a single administrative authority. These domains are separated due to differences in infrastructure, data ownership, policies, and operational constraints, each responsible for resource management and security. They may face unique complexity and scalability challenges while operating with varying infrastructure capabilities, such as Radio Access Technologies or computing resources within the edge-cloud continuum.

Within a single administrative domain, a MANO platform orchestrates network functions and other services in a federated (vertical or horizontal) fashion, interacting with the domain components and external consumers of the northbound/southbound APIs. This becomes particularly complex when multiple administrative domains need to be orchestrated in parallel (e.g., as in~\cite{5gvios}) or when multiple network segments or ``edges'' (each potentially with its own orchestrator) are part of the same administrative domain (e.g. as in \cite{10597106}).

Based on the above, we present Cloud-native Autonomous Management and INtent-based Orchestrator (CAMINO), a scalable MANO architecture that enables the orchestration of services within multi-level administrative domains. We build upon traditional NFV MANO principles and extend them across a modernised architecture that can accommodate the requirements of a 6G and cloud-native system. Compared to other platforms, it facilitates a configuration-centric approach of zero-touch provisioning of deployment intents by enabling closed-loop multi-edge orchestration featuring service brokering, resource monitoring and admission control capabilities. In this paper, we focus on the architecture of CAMINO, the design decisions and the initial implementation of the idea, discussing the lessons learned, the challenges faced, and the enhancements provided compared to existing solutions.

The remainder of this paper is as follows. Sec.~\ref{sec:background} presents related work on MANO platforms, highlighting existing approaches and limitations. Sec.~\ref{sec:architecture} introduces the CAMINO architecture, detailing its functional components and their interactions. Sec.~\ref{sec:implementation} describes the implementation of CAMINO, including the tools and technologies used. Finally, Sec.~\ref{sec:conclusion} concludes the paper and outlines potential future directions.

\begin{figure*}[ht!]
    \centering
    \includegraphics[width=\linewidth]{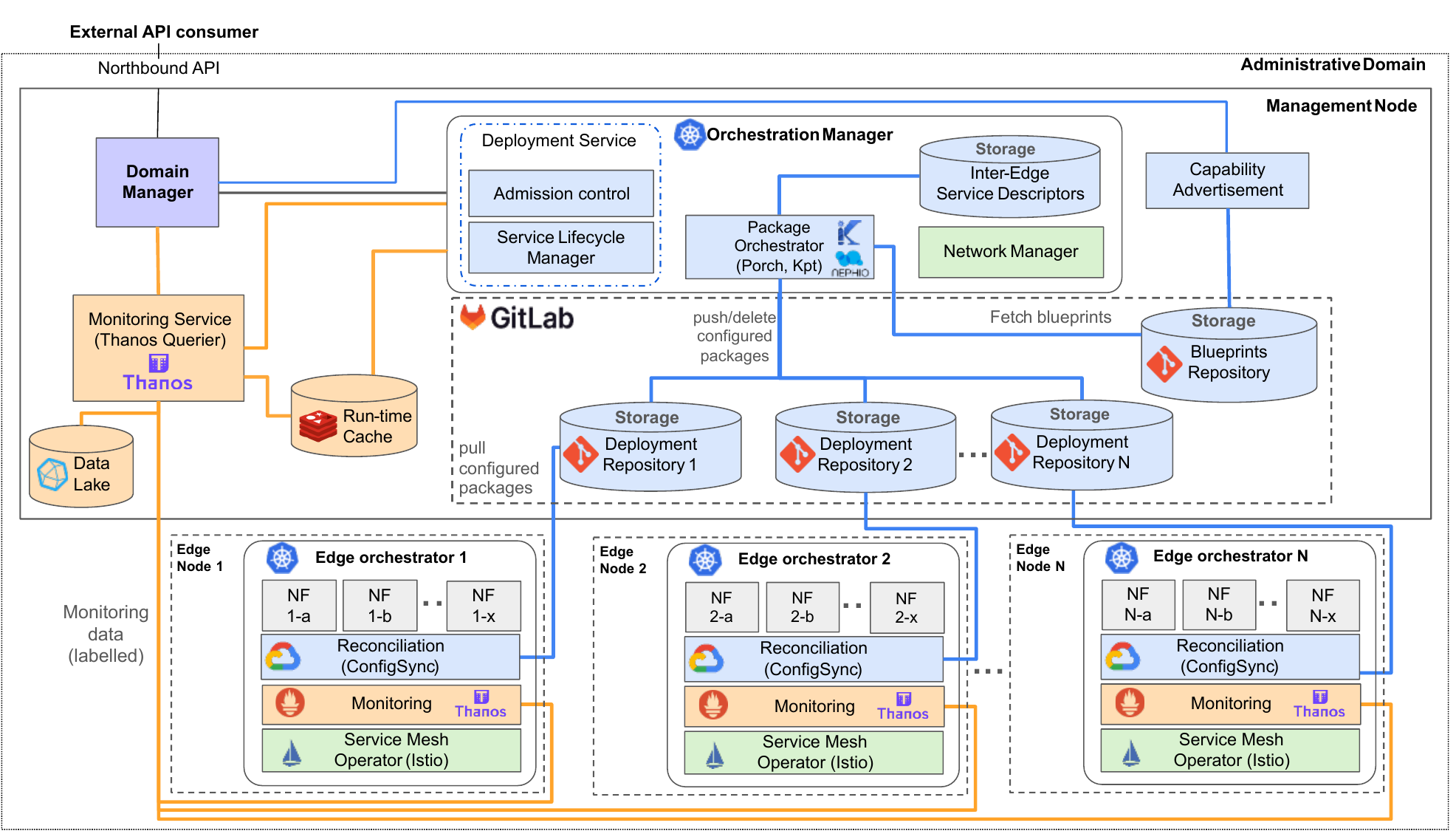}
    \vspace{-17pt} % Adds vertical space between image and caption
    \caption{Functional block diagram of CAMINO showing service orchestration (blue), monitoring (orange) and network management (green) elements.}
    \label{diagram}
    \vspace{-10pt} % Adds vertical space between image and caption
\end{figure*}

\vspace{-2mm}
\section{Related Work}
\label{sec:background}
%The increased complexity of cloud and edge deployments has led to the growing demand for efficient orchestration solutions capable of managing the workload lifecycle within a single administrative domain. Due to the heterogeneity and continuous evolution of infrastructure technologies, MANO platforms have been created to abstract and automate the management and configuration of such technologies.

The increased complexity of cloud and edge deployments has led to the creation of MANO platforms that abstract and automate the management and configuration of such technologies. In~\cite{10375939}, the survey underscores the value of a declarative approach to network slice provisioning with the translation of high-level service requirements into technical descriptions of network slices or subnets capable of fulfilling them. It is also noted that there is a lack of guidelines on how declarative provisioning APIs are designed. CAMINO builds upon this principle by decoupling the configuration values from the application code and the declarative deployment files, building upon the Configuration-as-Data (CaD) concept.

Two established open-source end-to-end (E2E) MANO platforms are the Open Network Automation Platform (ONAP)~\cite{onap_website} of Linux Foundation and the Open Source MANO (OSM)~\cite{osm_website} from ETSI. They both support cloud infrastructure platforms such as OpenStack and Kubernetes to deploy VNFs and CNFs. However, both solutions have inherent complexity from legacy telecom use cases and ETSI standards, and neither ONAP nor OSM supports a declarative way of service provisioning~\cite{10529648}. CAMINO brings a modular architecture, building upon declarative and intent-driven models and provides a solution that moves beyond traditional NFV, aiming to manage modern multi-domain-multi-edge network resources and services.

Another Linux Foundation project that follows the above principles is Nephio~\cite{nephio_website}, an intent-based cloud-native platform for the automated deployment and management of network functions based on CaD. However, Nephio is not an E2E MANO platform but acts as an abstraction layer between the E2E MANO platform layer and the layer of edge-level orchestrators/controllers. Furthermore, Nephio is a set of interchangeable open-source components per the CaD and intent-based orchestration principles. CAMINO utilises Nephio, leveraging how it handles configuration data and extends it with a more robust and scalable management and orchestration toolchain that allows for better lifecycle management (LCM) of service and network functions.

Even though the platforms above strive to provide holistic MANO solutions, they still fail to adequately address several critical administrative domain activities such as infrastructure management, SLA management and assurance, optimised placement and scaling, and policy enforcement, requiring additional components or extensions to bridge these gaps. The authors of~\cite{10529648} propose an architecture for zero-touch service MANO with run-time mechanisms that support it, but does not address scalability in VIM configuration management. They also focus on VM-based deployments such as OSM.

Authors in~\cite{eurecom_2024} highlight the importance of transitioning from VM-based to CNF-based workloads, introducing a Kubernetes Operator plane where multiple custom Kubernetes Operators declaratively implement MANO functionalities. However, this work does not focus on scalability and configuration management. 

Finally,~\cite{10575914} focuses on managing cloud-native infrastructures using a Logical Kubernetes Cluster and scheduling cloud-native workloads but without addressing the required data structures required to orchestrate the deployments and the coordination with other similar platforms. CAMINO holistically addresses the above issues and provides a uniform solution envisioning to modernise MANO solutions in future networking systems.

\vspace{-1mm}
\section{System architecture}
\label{sec:architecture}
Fig.~\ref{diagram} shows the functional blocks of the CAMINO architecture and their interconnections within the same administrative domain. The overall system enables the management and deployment of network function workloads in multiple \textit{Edge Orchestrators}, monitoring resources for evaluating running deployments and admission control of deployment intents.

The edge orchestrators are controlled by the \textit{Orchestration Manager}. We consider cloud-native Edge Orchestrators as well as a cloud-native Management node. For service and network deployments across administrative domains, a cross-domain orchestrator (CDO) is required. The following sections describe CAMINO's function blocks, highlighting configuration details, design decisions, and relevant interconnections.

\begin{lstlisting}[language=json,caption=An example service deployment intent (JSON) on a single administrative domain (Domain-X) considering a network function chain with a CNF placed in a different administrative domain (Domain-Y).,label=lst:deploy]
{
  "domain_name": "Domain-X",
  "deployment_id": "338d10a2-2669-46e1",
  "timestamp": "2025-01-24T20:55:50.991211",
  "services": [
    {
      "package_name": "CNF-1",
      "version": "v1",
      "qos_level": "default"
    },{
      "package_name": "CNF-2",
      "version": "v3",
      "qos_level": "default",
      "dependencies": [
        {
          "after": "CNF-3",
          "domain": "Domain-Y",
          "fqdn": "yyy.yyy.yyy.yyy"
        },{
          "after": "CNF-1",
          "domain": "Domain-X",
          "fqdn": "xxx.xxx.xxx.xxx"
        }
      ]
    },{
      "package_name": "CNF-4",
      "version": "v2",
      "qos_level": "default",
      "dependencies": [
        {
          "after": "CNF-2",
          "domain": "Domain-X",
          "fqdn": "xxx.xxx.xxx.xxx"
        }
      ]
    }
  ]
}
\end{lstlisting}

\vspace{-2mm}

\subsection{Deployment Intents and Packages}
A deployment intent represents an E2E service request deployed across multiple edge clusters (e.g., due to latency or privacy constraints of the E2E application). Listing~\ref{lst:deploy} provides an example deployment intent. In our architecture, this E2E deployment intent comprises one or multiple service components, referred to as \textit{packages}. These may include VNFs, CNFs, or other cloud-native components and configurations that collectively fulfil specific network and computing tasks. For example, a \textit{package} may be an Nginx ingress controller, while another may define an Ingress rule (e.g., in a Kubernetes cluster). A deployment intent is a high-level description of all the services and network configurations, including their associated package names, relevant resource requirements and dependencies. 

To manage deployment intents efficiently, we adopt the CaD principle, treating packages as \textit{declarative configuration bundles} managed by version control systems~\cite{9678525}. This allows centralised and simplified tracking of changes, package versioning and collaborative authoring, with the benefits of a declarative approach that separates configurations from the code that operates them. The configurations are divided into blueprint ("dry") configurations, which contain the component assembly information, and deployment ("hydrated") configurations, that is, the declarative statement of the desired state. Consequently, we classify the repositories that host packages into \textit{blueprint repositories} and \textit{deployment repositories}, respectively.

\begin{figure}
    \centering
    \includegraphics[width=0.8\linewidth]{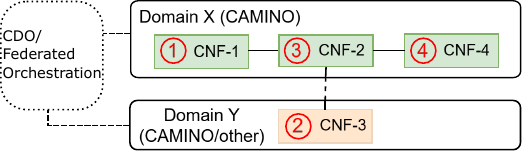}
    \vspace{-6pt} % Adds vertical space between image and caption
    \caption{Example deployment of a network function chain based on  Listing~\ref{lst:deploy}, and the order of deployment (in red), between two administrative domains.}
    \label{fig:deployment}
    \vspace{-12pt} % Adds vertical space between image and caption
\end{figure}

\subsection{Order of Deployment and Service Brokering}
Listing~\ref{lst:deploy} shows the deployment order and interdependencies across different network and service functions. The order and interdependencies can either be chosen by an end-user or automatically generated as part of a service brokering. For example, such a deployment intent can originate from a user portal or a cross-domain orchestrator (as in~\cite{reason_architecture}). 

In Fig.~\ref{fig:deployment}, and according to Listing~\ref{lst:deploy}, we see the execution order for four CNFs, three deployed by CAMINO in the administrative Domain X and one handled by a different administrative authority in Domain Y. The order dictates the generation of the required deployment configuration files for the entire chain. This is crucial because specific configurations (e.g., a Kubernetes ``service'', or the FQDN of a different domain) must be used as inputs to the following packages. Since JSON captures associations without order, we adopt a "labelling" approach, using the "after" label to control the deployment sequence. Based on the given labels, the Domain Manager performs a basic topological ordering and constructs a directed graph of the services with linear ordering. This graph represents the order of execution.

\subsection{Reconciliation and Deployment Configuration}
Based on the above order, the blueprint configurations describing a package are modified to compile a \textit{deployment configuration}. These configurations are manifests stored in the corresponding deployment repositories. Each repository is linked to a \textit{reconciliation operator} installed in each edge node (e.g., Kubernetes cluster). This operator synchronises the state of an edge deployment with the contents of its assigned deployment repository, which serves as the source of truth. In principle, when a configuration is added to or updated in a repository, the edge orchestrator automatically deploys it. Similarly, when a configuration is removed, the service is terminated.

To automate the declarative distribution and customisation of one or more blueprint packages, an Inter-Edge Service Descriptor (IESD) is required, stored in a separate storage entity. The IESD bundles the minimum deployment requirements and additional configurations to compile new manifests. All blueprint configurations include various parameterised field values and act as templates, enabling CRUD operations during deployment without violating the declarative manifest's schema.

For example, if an intent requires a package to be deployed, a JSON document as in Listing~\ref{lst:package} is created and used to generate the deployment configuration. If the deployment corresponds to a Kubernetes Deployment resource, the ``version'' and ``package\_name''  labels can generate the container image tag, the resources can be used as part of the Kubernetes resource request or limits (i.e., in \textit{spec.template.spec.containers[].resources}), while the ``qos'' label can control an environmental variable that modifies a set of exposed QoS configurations from the application. Additionally, different labels can be included (e.g., network resources, network slices, specific naming conventions, etc.) depending on application requirements. 

\vspace{-1mm}
\begin{lstlisting}[language=json,caption=Description of a package instance in JSON format, label=lst:package]
{
  "name": "example_package",
  "package_requirements": [{
    "qos": "default",
    "revision": "v5",
    "package_resources": {
      "container": "example_container",
      "cpu": 8,
      "memory": "1000000Ki"
    }
  }]
}
\end{lstlisting}
\vspace{-2mm}

\subsection{Orchestration Manager}

Each of the packages stored in the blueprints repository is a potential service or configuration that can be deployed by the Orchestration Manager once a deployment intent is forwarded by the Domain Manager. The Orchestration Manager is a logical entity composed of subcomponents that provide the following capabilities:

\begin{itemize}
    \item Registration/deletion of blueprint and deployment repositories.
    \item Service LCM and distribution of configuration bundles from blueprint repositories to deployment repositories.
    \item Manipulation of blueprint packages, including composition or selection of the appropriate IESDs.
    \item Configuration of network resources consumed by services of each deployment.
    \item Discovery of minimum requirements for selected intents to enable admission control using the monitoring data.
\end{itemize}

\subsection{Admission Control and Resource monitoring}
The role of the Admission Controller is twofold. Initially, it checks the configurations and IESDs (validates the schema or the content of the files) to prevent misconfiguration. Moreover, it verifies whether the required resources are available for a service (e.g., enough CPU cores and RAM) or a configuration (e.g., an Ingress Controller is available). 

A monitoring plane is required to track resource utilisation. The monitoring tools at the edge clusters collect the available resources and monitor the existing deployments. The monitoring service at the management node ensures enough resources are available at each cluster, raises alerts for misbehaving applications and aggregates the monitoring data for visualisation purposes.

There are two kinds of resources being monitored:
\begin{enumerate*}
    \item \textbf{Infrastructure resources}: overall metrics of each cluster within the administrative domain.
    \item \textbf{Workload resources}: metrics specific to each deployed network function.
\end{enumerate*}

External consumers (e.g., a cross-domain orchestrator~\cite{reason_architecture}) can also request access to monitoring data through the Domain Manager APIs. These requests are subject to administrative policies (e.g., providing only aggregated infrastructure utilisation instead of edge-level information). The Admission Control and Service Lifecycle Manager are responsible for deploying the configurations across the various edge nodes.

\begin{lstlisting}[language=json,caption=Description of a network deployment intent extracted from Listing~\ref{lst:deploy}, label=lst:network]
{
  "deployment_id": "338d10a2-2669-46e1",
  "services": [
  {
    "name": "CNF-1",
    "endpoints": [{
      "host": "svc1", "port": 80, "protocol": "HTTP"
  }],
    "links_to": [{
      "name": "CNF-2", "type": "intra-edge"
    }]
  },
  {
    "name": "CNF-2",
    "endpoints": [{
      "host": "svc2", "port": 80, "protocol": "HTTP"
    }],
    "links_to": [{
      "name": "CNF-1","type": "intra-edge"
    },{
      "name": "CNF-3",
      "type": "cross-domain",
      "resolution": {
        "domain": "Domain-Y", "fqdn": "yyy.yyy.yyy.yyy"
      }
    },{
      "name": "CNF-4", "type": "inter-edge"
    }]
  },
  {
    "name": "CNF-4",
    "endpoints": [{
      "host": "svc4", "port": 80, "protocol":"HTTP"
    }],
    "links_to": [{
     "name": "CNF-2", "type": "inter-edge"
    }]
  }]
}
\end{lstlisting}
\vspace{-1mm}

\subsection{Domain Manager}
The Domain Manager serves as the only point of contact for external entities. It provides the following capabilities using its northbound API endpoints: 

\begin{itemize}
     \item Advertises the available services and configurations of the blueprint repository as well as the total reserved resources to trusted external users or higher-level orchestrators.
     \item Receives deployment and termination intents, translating the relevant segments of these requests to the APIs of the Orchestration Manager's service LCM and network management components.
     \item Creates the topological linear ordering for the deployment.
     \item Manages authorisation and authentication for northbound endpoint calls.
     \item Discovers external domain network information (e.g., FQDN) in cooperation with other Domain Managers.
     \item Queries monitoring data from the Monitoring service.
     \item Responds to external health-check requests.
\end{itemize}

\subsection{Network Manager}
\label{sec:net-mgr}

% The Network Manager is an entity within the Orchestration Manager, responsible for enabling connectivity among services. This process is initiated after it receives connectivity configuration details extracted from the deployment intent. Such information can be mandatory, such as the topology, or optional such as policy enforcement, failover mechanisms or security requirements.
The Network Manager entity enables a programmable communication fabric across services. Establishing the fabric begins after the Network Manager receives the connectivity configuration details extracted from the deployment intent. Such configuration may include topology directives, policy enforcement rules, failover mechanisms, and security constraints. 

% To ensure interconnectivity among the deployed services, CAMINO leverages the service mesh paradigm, primarily as an effective solution for inter-edge deployments.
%an effective solution for robust inter-edge deployments.

% Furthermore, apart from simply enabling interconnectivity, service meshes act as abstraction layers that offer enhanced security, traffic management, observability, and policy enforcement across multiple Edge Node clusters. Of course, this is also true in the case of interconnecting services deployed within different Administrative Domains (that might or might not use the CAMINO architecture).
CAMINO leverages the Service Mesh paradigm, which abstracts inter-service communication by providing a control plane that accepts high-level declarative definitions, translating them into low-level network configuration, and injecting them into a data plane that spans the target services. Such abstraction simplifies various connectivity configurations such as:
\begin{enumerate*}
    \item \textbf{Intra-edge}: The minimum required connectivity configuration within a single cluster.
    \item \textbf{Inter-edge}: Connectivity across multiple clusters within the same network domain.
    \item \textbf{Cross-domain}: Connectivity across different administrative domains using CAMINO or a compatible solution.
\end{enumerate*}

Upon receiving the deployment intent, the Domain Manager extracts the deployment information (i.e., the associated packages and topology) and forwards it to the Orchestration Manager. Then, the Network Manager generates and implements a service mesh configuration based on the brokering plan that distributes the services to Edge nodes. An example is found in Listing~\ref{lst:network}. This information is used to hydrate blueprint network configuration files stored in the edge deployment repositories and deployed in the edge clusters.

\begin{figure}
    \centering
    \includegraphics[width=0.9\linewidth]{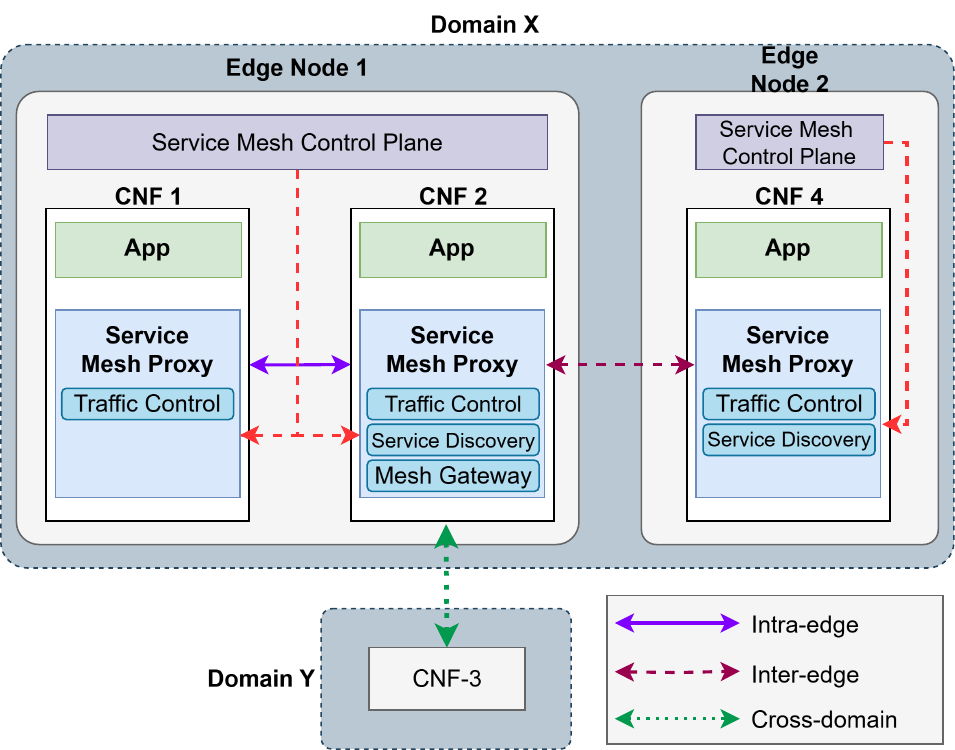}
    \vspace{-7pt} % Adds vertical space between image and caption
    \caption{Service Mesh deployment (as in Listing~\ref{lst:deploy}) of a network function chain placed between two different administrative domains. }
    \label{fig:servicemesh}
    \vspace{-17pt} % Adds vertical space between image and caption

\end{figure}

Fig.~\ref{fig:servicemesh} illustrates an example service mesh deployment based on the deployment intent of Listing~\ref{lst:deploy}. Each Edge Node has its own Service Mesh Control Plane that injects and configures a Service Mesh Proxy in each deployed CNF. For this example, we assume that in Domain X, the Orchestration Manager assigns CNFs 1 and 2 to Edge Node 1, so the Network Manager needs to establish intra-edge communication between the two CNFs. CNF 4 is assigned to Edge Node 2; thus, inter-edge communication between CNF2 and CNF4 is required. Domain Y hosts CNF3, so cross-domain communication between CNF2 and CNF3 is necessary. For cross-domain discovery, the Network Manager can either receive the remote domain FQDN from an end-user, request this information from a trusted cross-domain orchestrator, or maintain a list of trusted domains within the Domain Manager.

\vspace{-1mm}

\section{Implementation}
\label{sec:implementation}
This section describes the tools and solutions used or developed that enable E2E service deployments. In our setup, all nodes -both management and edge- are considered to be Kubernetes clusters with a MetalLB load-balancer, Calico as a Container Network Interface, and Nginx as the Ingress Controller. It is important to note that other solutions are acceptable, provided they comply with CAMINO's architecture. 

\subsection{Orchestration and Domain Management}
The package repositories are managed by GitLab, chosen due to its open-source nature and privacy preservation (i.e., local deployment). Our package manager of choice is \textit{kpt}~\cite{kpt_website}, used for automated authoring and manipulation of Kubernetes Resource Model (KRM) packages, e.g. \cite{example_packages}. Kpt was selected over Helm as it natively integrates with CaD and GitOps workflows. Additionally, it treats upstream packages as immutable artefacts with ``pull'' and ``customisation'' operations being separate steps. This prevents local changes from being overwritten, making updates easier to manage.  

Kpt introduces functions (e.g., setters) that enable CRUD operations on blueprint packages using IESD descriptors. All blueprint packages are described with parameterised, customisable fields in the form of comments within the defined YAML manifest. These comments dictate the parameters that can be manipulated such as metadata (e.g., namespaces, names, labels), spec (e.g., images, resources, selectors) and others. All parameters are managed with IESDs in the form of PackageVariantSets (PVSs), a custom resource (CR) managed by \textit{porch}~\cite{nephio_website}. The blueprint packages are reusable and can become deployment packages through ``hydration''. In general, \textit{porch} handles creating, deleting, and maintaining the blueprint and deployment package repositories within GitLab, registering all repositories across management and edge nodes, and generating and deploying porch-compliant CRDs. 

The \textit{Package Orchestrator} (developed in Python 3.11) wraps around kpt and \textit{porch}, automating the creation and management of PVSs, kpt function manipulation, and the storage or retrieval of all packages. The \textit{Domain Manager} (developed in Python 3.11) handles creating, deleting, and maintaining the blueprint and deployment package repositories within GitLab and triggers the registration of new repositories across all management and edge clusters. Upon receiving a deployment intent and generating a topological order, the intent JSON is sent to the \textit{Deployment Service} (developed in Python 3.11). This service combines the \textit{Service Lifecycle Manager} and the \textit{Admission Controller} functionality. It handles package deployment proposals and approves them under certain conditions (e.g., if no misconfiguration exists, or if enough resources are available for a deployment). Misconfigurations are checked by ``dry-running'' the updated/hydrated packages and evaluating the response of the edge nodes. Resources are queried from the \textit{Monitoring Service} (both workload and infrastructure resources) or directly from the Kubernetes API if it concerns cluster or network configuration resources. 

A package approval triggers \textit{porch} to push the package to a deployment repository. Later, the \textit{Reconciliation Manager} of each edge node handles the package deployment using the local orchestrator. Similarly, when a termination intent is received, the \textit{Deployment Service} checks whether it is possible or if a conflict may arise and accordingly handles the termination, instructing \textit{porch} to delete the deployment package, and the reconciliator terminates the Kubernetes deployment. As a reconciliation tool, ConfigSync \cite{configsync_github} was chosen due to its edge-based architecture which supports horizontal scaling in the case of a large number of edge nodes, compared to other solutions, such as ArgosCD, that follow a server-client model.

\vspace{-1mm}
\subsection{Network Management}
% As previously mentioned, the role of the Network Manager is to obtain the network topology of the submitted service deployment and generate the appropriate configuration of a service mesh. In our case, the network management service is leveraging Istio, an open-source service mesh that provides policy-driven traffic management, security, and observability features for distributed applications.

The \textit{Network Manager} (developed in Python 3.11) implements the Service Mesh paradigm leveraging Istio Service Mesh~\cite{istio_website}. Once a deployment is triggered by the \textit{Deployment Service}, and if communication with another deployment is required, an Istio Envoy Proxy Sidecar is injected in each Pod. This enables traffic management by intercepting traffic and applying policy-based routing. 

The \textit{Network Manager} is responsible for generating all Istio CRDs, adhering to CaD principles and utilising pre-existing network blueprint packages. The CRDs created are similarly deployed alongside the Pods across all edge nodes and are translated by the Istio control plane into low-level Proxy configurations. To provide isolation across deployments, we group CNFs of a specific chain into Kubernetes Namespaces and create and apply network policies using and adjusting pre-existing blueprint packages. Finally, we consider three different connectivity configurations:
\begin{enumerate*}
    \item \textbf{Intra-edge}: Leveraging the Virtual Service (VS) and Destination Rule (DR) CRDs to implement network policies, such as load balancing.
    \item \textbf{Inter-edge}: Istio's model for individual multi-cluster control planes simplifies inter-edge service discovery and enables high availability using the VS, DR and Service Entry CRDs for remote service reachability. Additionally, we follow the \textit{Namespace Sameness} principle to identify remote services.
    \item \textbf{Cross-Domain (External)}: Cross-network Control Planes enable remote service discovery via the East-West Gateway CRD, allowing clusters to expose local services on specific ports through the domain's External IP.
\end{enumerate*}

\vspace{-1mm}
\subsection{Monitoring}
Finally, the monitoring capabilities are enabled by Thanos and Prometheus~\cite{thanos_website} instances in each Edge Node to capture performance data for the Edge Nodes and the services. Each Prometheus instance in each cluster hosts a Thanos agent as a sidecar. All agents are queried by a Thanos Querier instance controlled by the management cluster's \textit{ Monitoring Service} (developed in Python 3.11). The \textit{ Monitoring Service} requests monitoring data for each metric of interest from all Thanos agents with a single PromQL query. This data can then be passed to relevant data-consuming services such as the Orchestration Manager or the Domain Manager. 

Monitoring data are separated with custom labels, which are assigned during the service deployment or cluster instantiation. For example, the Deployment service assigns pod names a unique identifier, and edge nodes have unique contexts and names. The labels are exchanged between the Orchestration Manager and the Monitoring Service using a run-time cache (Redis was chosen for its lightweight nature and stability). Finally, all monitoring data is stored in a data lake (using InfluxDB) for long-term analysis and visualisation.

\vspace{-1mm}
\section{Conclusion}
\label{sec:conclusion}

This paper introduces CAMINO, a cloud-native, autonomous management and intent-based orchestration framework designed to address the complexity of service deployment across multiple edge nodes leveraging the CaD principle. Its service orchestration, network management and monitoring components enable scalable, zero-touch provisioning of heterogeneous services to meet the demands of 6G systems. The CAMINO framework enhances the efficiency and observability of complex edge-cloud deployments. Future work will extend its intent-based capabilities with AI-driven service management and cross-administrative domain orchestration. This research contributes to the advancement of autonomous network orchestration, paving the way for scalable, resilient, and self-managing telecommunications infrastructures.

\vspace{-1mm}
\section*{Acknowledgements}
This work is a contribution by Project REASON, a UK Government funded project under the Future Open Networks Research Challenge (FONRC) sponsored by the Department of Science Innovation and Technology (DSIT).

\vspace{-1mm}
\bibliographystyle{IEEEtran}
\bibliography{bibliography}

\end{document}